\documentclass[a4paper,english,titlepage,12pt]{article}

\usepackage{babel}\usepackage{amsmath,amssymb,amsthm}
\usepackage[numbers,compress]{natbib}
\usepackage{graphicx}
\usepackage{ctable}
\usepackage{float}

\newcommand{\new}{\text{\textit{\textbf{new}}}}
\newcommand{\rop}{\mathcal{R}}

\begin{document}

\begin{center}\LARGE{\textbf{Subtraction Procedure for Calculation
of Anomalous Magnetic Moment of Electron in QED and its Application
to Numerical Computation at 3-loop Level}}
\end{center}
\begin{center}
\large{S. A. Volkov\footnote{E-mail: \texttt{volkoff\underline{
}sergey@mail.ru}}}
\\ \emph{\small{SINP MSU, Moscow, Russia}}
\end{center}

\small{A new subtraction procedure for removal both
ultraviolet and infrared divergences in Feynman integrals is proposed.
This method is developed for computation of QED corrections to the electron
anomalous magnetic moment. The procedure is formulated in the
form of a forest formula with linear operators that are applied to
Feynman amplitudes of UV-divergent subgraphs. The contribution of
each Feynman graph that contains propagators of electrons and
photons is represented as a finite Feynman-parametric integral.
Application of the developed method to the calculation of
2-loop and 3-loop contributions is described.} \normalsize

\section{Introduction}

The Bogoliubov-Parasiuk theorem ~\cite{bogolubovparasuk,hepp} provides
us a constructive definition of the procedure (R-operation) that
removes all ultraviolet divergences in each Feynman graph. The proof
of this theorem in ~\cite{bogolubovparasuk,hepp} gives the
representation of the Feynman amplitude that is obtained by
R-operation in the form of an absolutely convergent
Schwinger-parametric integral if the imaginary addition
$i\varepsilon$ ($\varepsilon>0$) to the propagator denominators is
fixed. Thus, R-operation removes UV-divergences point-by-point,
before integration. The explicit formula for R-operation was
obtained in ~\cite{scherbina,zavialovstepanov}:
\begin{equation}\label{eq_zavialovstepanov}
\rop=(1-M_1)(1-M_2)\ldots (1-M_n),
\end{equation}
where $M_j$ is the operator that extracts Taylor expansion of the
Feynman amplitude of $j$-th divergent subgraph up to the needed order
around zero momenta. Here it is meant that we should remove brackets
and delete all terms containing $M_j$ and $M_l$ that correspond to
overlapping\footnote{Subgraphs are said to overlap if their sets of
lines have non-empty intersection, and they are not contained one
inside the other.} subgraphs. In the same papers it was pointed out
that this renormalized Feynman amplitude can be represented in the
form of an absolutely convergent Schwinger-parametric integral when
$\varepsilon>0$ is fixed. Later, this formula was independently
rediscovered by Zimmermann in momentum representation
~\cite{zimmerman}, see also~\cite{zavialov,smirnov}.

Note that R-operation doesn't remove infrared divergences. For
example, in QED, if we consider external momenta in Feynman graphs
on the mass shell, then the Feynman amplitude that is renormalized
by R-operation doesn't converge to a distribution as
$\varepsilon\rightarrow 0$. Also, the physical renormalization
requires to take the on-shell renormalization operators instead of
$M_j$  in (\ref{eq_zavialovstepanov}), and these operators generate
additional IR-divergences (see~\cite{ll4,bogoliubov_shirkov}).

In this paper we consider a development of the R-operation idea.
This development is applied to the problem of calculation of QED
corrections to the electron's anomalous magnetic moment (AMM).

Electron's AMM is known with a very high accuracy, in the
experiment~\cite{experiment} the value
$$
a_e=0.00115965218073(28)
$$
(in Dirac moment units) was obtained. So, a maximal possible precision is
needed also from theoretical predictions. For high-precision calculations
it is required to take into account Feynman graphs with a large number of
independent loops, this requires a lot of computer resources.
Therefore, the ability to remove all divergences (including the
infrared ones) point-by-point is certainly relevant. Generally
speaking, electron's AMM in QED is free from infrared divergences in
each order of the perturbation series since IR-divergences of the unrenormalized
Feynman amplitude are cancelled by IR-divergences in renormalization
constants (about IR-divergences in renormalization constants, see
~\cite{ll4,bogoliubov_shirkov}). However, individual graphs remain
IR-divergent. Unfortunately, the structure of infrared and
ultraviolet divergences in individual graphs is complicated, the
divergences of different types can, in a certain sense, be
''entangled'' with each other. At the given moment, there is no any
universal method for subtraction of IR-divergences in QED Feynman
graphs.

The most accurate prediction of electron's AMM at the present
moment~\cite{kinoshita_10,kinoshita_10_new} has the following
representation:
$$
a_e=a_e(\text{QED})+a_e(\text{hadronic})+a_e(\text{electroweak}),
$$
$$
a_e(\text{QED})=\sum_{n\geq 1} \left(\frac{\alpha}{\pi}\right)^n
a_e^{2n},
$$
$$
a_e^{2n}=A_1^{(2n)}+A_2^{(2n)}(m_e/m_{\mu})+A_2^{(2n)}(m_e/m_{\tau})+A_3^{(2n)}(m_e/m_{\mu},m_e/m_{\tau}),
$$
where $m_e$, $m_{\mu}$, $m_{\tau}$ are masses of electron, muon, and
tau lepton, respectively. The value $A^{(2)}_1=0.5$ was obtained
analytically by Schwinger in 1948~\cite{schwinger1,schwinger2}. The
term $A_1^{(4)}$ was first calculated by Karplus and Kroll
~\cite{analyt2_kk} using combined numerical-analytical method but there
was a mistake in that calculation. This mistake was corrected
analytically by Petermann~\cite{analyt2_p} and Sommerfield~\cite{analyt2_z},
the new value was confirmed by using another
approach in~\cite{terentiev}:
\begin{equation}\label{eq_analyt2}
A_1^{(4)}=-0.328478965579193\ldots
\end{equation}
The value $A_1^{(6)}$ was computing numerically with the help of
computers by three groups of scientists in the first half of 1970s
(see ~\cite{levinewright}, \cite{carrollyao,carroll},
~\cite{kinoshita_6}). The most accurate value $A_1^{(6)}=1.195\pm
0.026$ for that period of time was obtained by Kinoshita and
Cvitanovi\'{c} ~\cite{kinoshita_6} (the error is due to the Monte
Carlo integration). By 1995, the accuracy was improved
~\cite{kinoshita_6_prec}: $A_1^{(6)}=1.181259(40)$. In all three
cases, the base of the method was a subtraction procedure for
point-by-point elimination of IR and UV divergences. These
subtraction procedures were developed especially for 3-loop
calculation of electron's AMM. However, in all three cases, a finite
renormalization after the subtraction was required. The rules for
this renormalization in the 3-loop case doesn't lead to an automated
procedure at any order of perturbation. Simultaneously,
approximately at the end of 1960s, the work of analytical
calculation of $A_1^{(6)}$ was started with help of computers (see
~\cite{analyt_mi,analyt_b2,analyt_b3,analyt_b1,analyt_b4,
analyt_b,analyt_e,analyt_d,analyt_c,analyt_ll,analyt_f,analyt3}
etc.). This work was finished in 1996 when in ~\cite{analyt3} the
value
\begin{equation}\label{eq_analyt3}
A_1^{(6)}=1.181241456\ldots
\end{equation}
was obtained. The first numerical values for $A_1^{(8)}$ were
computed by Kinoshita and Lindquist at the beginning of 1980s (see
~\cite{kinoshita_8start}), since then the accuracy is still
improving by Kinoshita and his collaborators. The numerical value
for $A_1^{(10)}$ was first obtained by Kinoshita's team in 2012
~\cite{kinoshita_10}. To realize that calculation, a new subtraction
procedure for point-by-point removal of divergences was developed.
The new method of Kinoshita and collaborators was fully automated up
to $A_1^{(8)}$, however, some individual Feynman graphs for
$A_1^{(10)}$ require a special treatment ~\cite{kinoshita_watanabe}.
In paper ~\cite{kinoshita_10_new} the recent results of computation
was presented:
$$
A_1^{(8)}=-1.91298(84),\quad A_1^{(10)}=7.795(336).
$$
The corresponding theoretical prediction
$$
a_e=0.001159652181643(25)(23)(16)(763)
$$
was obtained by using the value of the fine structure constant
$\alpha^{-1}=137.035999049(90)$ that had been measured in the recent
experiments with rubidium atoms (see ~\cite{rubidium,codata}). Here,
the first, second, third, and fourth uncertainties come from
$A_1^{(8)}$, $A_1^{(10)}$,
$a_e(\text{hadronic})+a_e(\text{electroweak})$ and the
fine-structure constant\footnote{Thus, the computed coefficients are
used for improving the accuracy of $\alpha$.} respectively. Let us
note that at the present moment there is no any independent check of
the calculation of $A_1^{(8)}$ and $A_1^{(10)}$, therefore, the
problem of computing $A_1^{(2n)}$ is still relevant. Some terms of
the expansion of $A_2^{(2n)}$ and $A_3^{(2n)}$ ($n\leq 4$) in powers
of $m_e/m_{\mu}$, $m_e/m_{\tau}$, logarithms of $m_e/m_{\mu}$ and
$m_e/m_{\tau}$ are known analytically (see
~\cite{analytheavy4_kataev,analytheavy4}). Also, these values and
$A_2^{(10)}(m_e/m_{\mu})$ were computed numerically (see
~\cite{kinoshita_10_new}), and the results of this computation for
$A_2^{(2n)}$, $A_3^{(2n)}$ ($n\leq 4$) are in good agreement with
the analytical ones.

In this paper we present a new subtraction procedure for calculation
of $A_1^{(2n)}$. This procedure eliminates IR and UV divergences
point-by-point, before integration, in the spirit of the papers
~\cite{bogolubovparasuk, hepp, scherbina, zavialovstepanov,
zavialov, smirnov, kinoshita_10, kinoshita_10_new, levinewright,
carrollyao, carroll, kinoshita_6} etc. The method has the following
advantages:
\begin{itemize}
\item The method is fully automated for any $n$.
\item The method is comparatively easy for realization on computers.
\item The given subtraction procedure is a modification of
(\ref{eq_zavialovstepanov}), it differs from that one only in the
choice of operators and in the way of combining them.
Operators of a simple form are used, which transform Feynman
amplitudes of subgraphs. The operators can be cast in the momentum
representation, they are linear, and produce polynomials of the degree that
is less or equal to the ultraviolet degree of divergence of the
corresponding subgraph\footnote{The subtraction procedures in
~\cite{kinoshita_10,levinewright,carrollyao,carroll,kinoshita_6} use
the operators that work with formulas in Feynman-parametric or
momenta representation (not with functions).}.
\item The contribution of each Feynman graph to $A_1^{(2n)}$ can be represented
as a single Feynman-parametric integral. The value of $A_1^{(2n)}$ is
a sum of these contributions. So, we don't need any additional
finite renormalizations, calculations of renormalization constants,
calculations of some values at the lower orders of perturbation, or
other additional calculations.
\item The given subtraction procedure was checked for 2-loop and
3-loop Feynman graphs by numerical integration. Most likely, it will
work at the higher orders of perturbation (the detailed explanation
is given in the full version of this paper~\cite{full_version}).
\item It is possible to use Feynman parameters directly. We don't
need any additional tricks\footnote{For example, one can use $m^2$
in propagators as additional variables of integration (see
~\cite{kinoshita_rules,kinoshita_6}).} to define Feynman parameters
in Feynman graphs that have non-negative UV degrees of divergence.
\end{itemize}
Presumably, the ideas of the given method can be applied to some other
problems.

\section{Formulation of the method}\label{sec_formulation}

\subsection{Preliminary remarks}

We will work in the system of units, in which $\hbar=c=1$, the
factors of $4\pi$ appear in the fine-structure constant:
$\alpha=e^2/(4\pi)$, the tensor $g_{\mu\nu}$ is defined by
$$
g_{\mu\nu}=g^{\mu\nu}=\left(\begin{matrix}1 & 0 & 0 & 0 \\ 0 & -1 &
0 & 0 \\ 0 & 0 & -1 & 0 \\ 0 & 0 & 0 & -1 \end{matrix}\right),
$$
the Dirac gamma-matrices satisfy the following condition
$\gamma^{\mu}\gamma^{\nu}+\gamma^{\nu}\gamma^{\mu}=2g^{\mu\nu}$.

We will use Feynman graphs with propagators
$\frac{i(\hat{p}+m)}{p^2-m^2+i\varepsilon}$ for electron lines and
\begin{equation}\label{eq_feynman_gauge}
\frac{-g_{\mu\nu}}{p^2+i\varepsilon}
\end{equation}
for photon lines. It is always assumed that a Feynman graph is
strongly connected and doesn't have odd electron cycles.

The number $\omega(G)=4-N_{\gamma}-\frac{3}{2}N_e$ is called the
\emph{ultraviolet degree of divergence} of the graph $G$. Here,
$N_{\gamma}$ is the number of external photon lines of $G$, $N_e$ is
the number of external electron lines of $G$.

If for some subgraph\footnote{In this paper we consider only such
subgraphs that are strongly connected and contain all lines that
join the vertexes of the given subgraph.} $G'$ of the graph $G$ the
inequality $\omega(G')\geq 0$ is satisfied, then UV-divergence can
appear. A graph $G'$ is called UV-divergent if $\omega(G')\geq 0$.
There are the following types of UV-divergent subgraphs in QED
Feynman graphs: \emph{electron self-energy subgraphs}
($N_e=2,N_{\gamma}=0$), \emph{photon self-energy subgraphs}
($N_e=0,N_{\gamma}=2$), \emph{vertex-like} subgraphs
($N_e=2,N_{\gamma}=1$), \emph{photon-photon scattering
subgraphs}\footnote{The divergences of this type are cancelled in
the sum of all Feynman graphs, but they can appear in individual
graphs.} ($N_e=0,N_{\gamma}=4$).

\subsection{Anomalous magnetic moment in terms of Feynman amplitudes}\label{subsec_feyn_ampl}

A set of subgraphs of a graph is called a \emph{forest} if any two
elements of this set are not overlapped.

For vertex-like graph $G$ by $\mathfrak{F}[G]$ we denote the set of
all forests $F$ containing UV-divergent subgraphs of $G$ and
satisfying the condition $G\in F$. By $\mathfrak{I}[G]$ we denote
the set of all vertex-like subgraphs $G'$ of $G$ such that each $G'$
contains the vertex that is incident\footnote{We say that a vertex
$v$ and a line $l$ are \emph{incident} if $v$ is one of endpoints of
$l$.} to the external photon line of $G$.\footnote{In particular,
$G\in \mathfrak{I}[G]$.}

Let us define the following linear operators that are applied to the
Feynman amplitudes of UV-divergent subgraphs:
\begin{enumerate}
\item $A$ --- projector of the anomalous magnetic moment. This
operator is applied to the Feynman amplitudes of vertex-like
subgraphs. Let $\Gamma_{\mu}(p,q)$ be the Feynman amplitude
corresponding to an electron of initial and final four-momenta
$p-q/2$, $p+q/2$. The Feynman amplitude $\Gamma_{\mu}$ can be
expressed in terms of three form-factors:
$$
\overline{u}_2 \Gamma_{\mu}(p,q) u_1 = \overline{u}_2 \left(
f(q^2)\gamma_{\mu} -\frac{1}{2m}g(q^2)\sigma_{\mu\nu}q^{\nu} +
h(q^2)q_{\mu} \right) u_1,
$$
where $(p-q/2)^2=(p+q/2)^2=m^2$,
$(\hat{p}-\hat{q}/2-m)u_1=\overline{u}_2(\hat{p}+\hat{q}/2-m)=0$,
$$
\sigma_{\mu\nu}=\frac{1}{2}(\gamma_{\mu}\gamma_{\nu}-\gamma_{\nu}\gamma_{\mu}),
$$
see, for example, \cite{ll4}. By definition, put
\begin{equation}\label{eq_a_def}
A \Gamma_{\mu} = \gamma_{\mu}\cdot \lim_{q^2\rightarrow 0}
(g(q^2)+C_Ah(q^2)),
\end{equation}
where $C_A$ is an arbitrary constant (the final result doesn't
depend on $C_A$, but contributions of individual Feynman graphs can
depend).
\item The definition of the operator $U$ depends on the type of
UV-divergent subgraph to which the operator is applied:
\begin{itemize}
\item If $\Pi$ is the Feynman amplitude that corresponds to a photon
self-energy subgraph or a photon-photon scattering subgraph, then, by
definition, $U\Pi$ is a Taylor expansion of $\Pi$ around zero
momenta up to the UV divergence degree of this subgraph.
\item If $\Sigma(p)$ is the Feynman amplitude that corresponds to an
electron self-energy subgraph,
\begin{equation}\label{eq_sigma_general}
\Sigma(p)=a(p^2)+b(p^2)\hat{p},
\end{equation}
then, by definition\footnote{Note that it differs from the standard
on-shell renormalization.},
$$
U\Sigma(p) = a(m^2)+b(m^2)\hat{p}.
$$
\item If $\Gamma_{\mu}(p,q)$ is the Feynman amplitude that
corresponds to vertex-like subgraph,
\begin{equation}\label{eq_gamma_general_q0}
\Gamma_{\mu}(p,0)=a(p^2)\gamma_{\mu} + b(p^2)p_{\mu} +
c(p^2)\hat{p}p_{\mu}+d(p^2)(\hat{p}\gamma_{\mu}-\gamma_{\mu}\hat{p}),
\end{equation}
then, by definition,
\begin{equation}\label{eq_u_vertex}
U\Gamma_{\mu}=(a(m^2)+C_Ud(m^2))\gamma_{\mu},
\end{equation}
where $C_U$ is an arbitrary constant.
\end{itemize}
\item $L$ is the operator that is used for on-shell renormalization
of vertex-like subgraphs. If $\Gamma_{\mu}(p,q)$ is the Feynman
amplitude that corresponds to a vertex-like subgraph,
$$
\Gamma_{\mu}(p,0)=a(p^2)\gamma_{\mu} + b(p^2)p_{\mu} +
c(p^2)\hat{p}p_{\mu}+d(p^2)(\hat{p}\gamma_{\mu}-\gamma_{\mu}\hat{p}),
$$
then, by definition,
\begin{equation}\label{eq_q_def}
L\Gamma_{\mu}=[a(m^2)+mb(m^2)+m^2c(m^2)]\gamma_{\mu}.
\end{equation}
\end{enumerate}

Let $f_G$ be the unrenormalized Feynman amplitude that corresponds
to a vertex-like graph $G$. By definition, put
\begin{equation}\label{eq_rop_tilde}
\tilde{f}_G=\rop^{\new}_G f_G,
\end{equation}
where
\begin{equation}\label{eq_rop}
\rop^{\new}_G=\sum_{\substack{F=\{G_1,\ldots,G_n\}\in \mathfrak{F}[G] \\
G'\in \mathfrak{I}[G]\cap
F}}(-1)^{n-1}M^{F,G'}_{G_1}M^{F,G'}_{G_2}\ldots M^{F,G'}_{G_n},
\end{equation}
\begin{equation}\label{eq_operators}
M^{F,G'}_{G''}=\begin{cases}A_{G'},\text{ if }G'=G'', \\
U_{G''},\text{ if }G''\notin \mathfrak{I}[G]\text{ or }G''\subseteq
G', G''\neq G',
\\ L_{G''},\text{ if }G''\in \mathfrak{I}[G], G'\subseteq G'', G''\neq
G, G''\neq G',
\\ (L_{G''}-U_{G''}),\text{ if }G''=G, G'\neq G.\end{cases}
\end{equation}
In this notation, the subscript of a given operator symbol denotes the
subgraph to which this operator is applied.

By $\check{f}_G$ we denote the coefficient before $\gamma_{\mu}$ in
$\tilde{f}_G$. The value $\check{f}_G$ is the contribution of
graph $G$ to the anomalous magnetic moment:
$$
a_{e,1}^{\new}=\sum_{G} \check{f}_G,
$$
where the summation goes over all vertex-like Feynman graphs. If we sum
only over graphs with a fixed number of vertices, we can obtain
the corresponding term of the perturbation expansion in $\alpha$.

\begin{figure}
\includegraphics{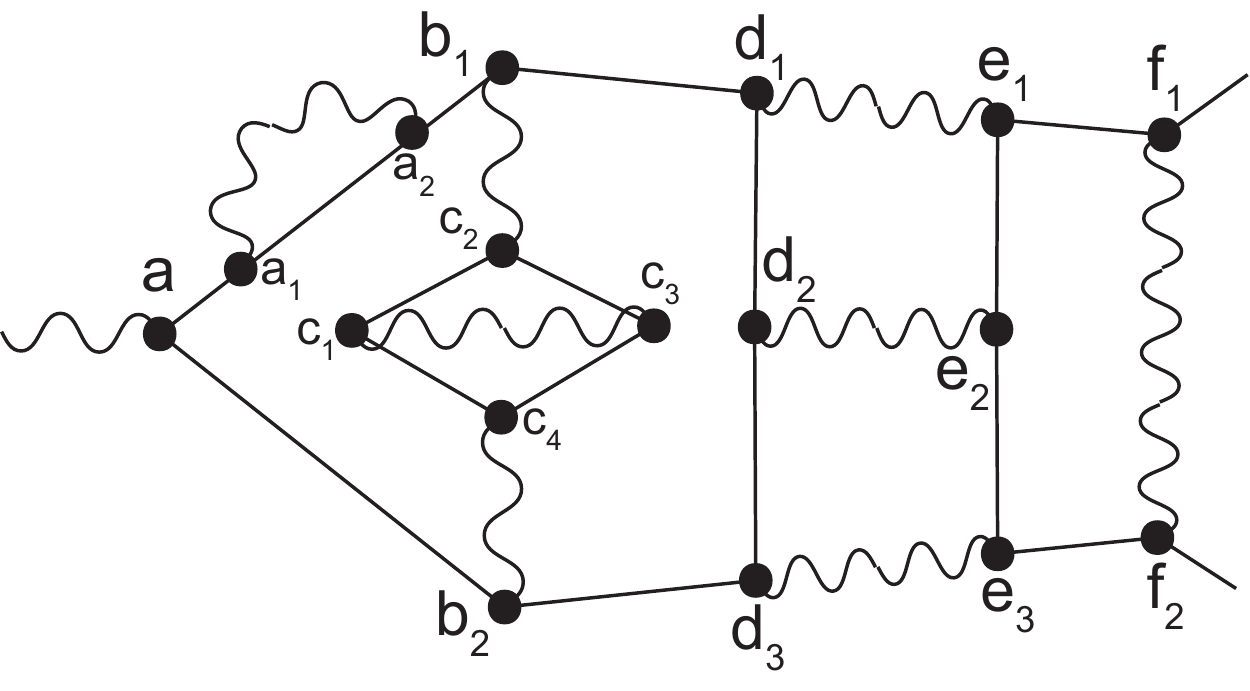}
\caption{A complicated Feynman graph (an example)}
\label{fig_example}
\end{figure}

Let us consider the example that is showed on Figure
\ref{fig_example}. This Feynman graph we denote by $G$.
For this example, we have
$$
\mathfrak{I}[G]=\{G_c,G_e,G\},
$$
where
$$
G_c=aa_1a_2b_1b_2c_1c_2c_3c_4,\quad
G_e=aa_1a_2b_1b_2c_1c_2c_3c_4d_1d_2d_3e_1e_2e_3
$$
(subgraphs are specified by enumeration of vertices). Also, there
are following UV-divergent subgraphs: $a_1a_2$ (electron
self-energy), $c_1c_2c_3$, $c_1c_3c_4$ (vertex-like, overlapping),
$c_1c_2c_3c_4$ (photon self-energy),
$G_d=aa_1a_2b_1b_2c_1c_2c_3c_4d_1d_2d_3$ (photon-photon scattering).
Using (\ref{eq_rop_tilde}), (\ref{eq_rop}), (\ref{eq_operators}) we
obtain
\begin{eqnarray}
\tilde{f}_G &=& \left[ A_G (1-U_{G_e})(1-U_{G_d})(1-U_{G_c}) -
(L_G-U_G)A_{G_e}(1-U_{G_d})(1-U_{G_c})  \right.
\nonumber \\
&-& \left. (L_G-U_G) (1-L_{G_e})(1-U_{G_d}) A_{G_c} \right]
\nonumber \\ \nonumber
&\times&
(1-U_{c_1c_2c_3c_4})(1-U_{c_1c_2c_3}-U_{c_1c_3c_4})(1-U_{a_1a_2})
f_G.
\end{eqnarray}

Operator expressions for 2-loop Feynman graphs are given in Table
\ref{table_2loop} (in this table by $G$ we denote the whole graph).

\subsection{Feynman-parametric representation}\label{subsec_parametric}

Let us consider the formulation of the subtraction procedure in
Feynman-parametric representation. This representation allows to
remove regularization, so it can be directly used for numerical
calculation.

We will use the following formula:
$$
\frac{1}{x+i\varepsilon}=\frac{1}{i}\int_0^{+\infty}
e^{iz(x+i\varepsilon)}dz.
$$

To calculate the contribution of vertex-like graph $G$ to
$a_{e,1}^{\new}$ in terms of Feynman parameters we should perform
the following steps:
\begin{enumerate}
\item To each internal line of $G$ we assign the variable $z_j$,
where $j$ is the number of this line.
\item Suppose that the values $z_j>0$ are fixed. We introduce the
following propagators for electron and photon lines respectively:
\begin{equation}\label{eq_exp_propagators}
(\hat{p}+m)e^{iz_j(p^2-m^2+i\varepsilon)},\quad ig_{\mu\nu}
e^{iz_j(p^2+i\varepsilon)}.
\end{equation}
By $\check{f}_G(\underline{z},\varepsilon)$, where
$\underline{z}=(z_1,z_2,\ldots)$, we denote the value that is
obtained by the rules that are described above for $\check{f}_G$,
but with the use of the new propagators. The value
$\check{f}_G(\underline{z},\varepsilon)$ is obtained by using
explicit formulas for integrals of multi-dimensional gaussian
functions multiplied by polynomials, see~\cite{zavialov,smirnov}.
After applying these explicit formulas to a Feynman integral with
propagators (\ref{eq_exp_propagators}) we obtain the Feynman
amplitude of the form
\begin{equation}\label{eq_ampl_z}
\Pi_1 R_1 + \Pi_2 R_2 + \ldots + \Pi_N R_N,
\end{equation}
where each $\Pi_l$ can be represented as a product of expressions
like $\hat{p}_j$, $p_{j\mu}$, $\gamma_{\mu}$, $(p_{j'}p_{j''})$
(here $p_1,p_2,\ldots$ are momenta of external lines, $j$, $j'$,
$j''$ are coordinate indexes of external momenta, $\mu$ is the
tensor index that corresponds to the external photon line), each
$R_l$ has the form
$$
\frac{F(\underline{z})}{T(\underline{z})}\cdot \exp
\left[i\frac{H(\underline{p},\underline{z})}{T(\underline{z})}-\varepsilon\sum
z_j\right],
$$
where $\underline{p}=(p_1,p_2,\ldots)$ is the tuple of external
momenta, $F$, $T$, $H$ are homogeneous polynomials with respect to
$z$, all coefficients of the polynomial $T$ are positive, all
coefficients of $H$ are real, the degree of $H$ with respect to $z$
is equal to $1$ plus the degree of $T$, the polynomial $H$ contains
elements of $\underline{p}$ only in the form of scalar products like
$(p_{j'}p_{j''})$, and each term of $H$ contains not more than one
such scalar product (see
~\cite{zavialov,smirnov,bogoliubov_shirkov}).

If operators $A$, $U$, and $L$ are applied to some subgraphs of a
given Feynman graph, then the corresponding Feynman amplitude can be
represented in the form (\ref{eq_ampl_z}) too. This can be proved by
induction on the number of internal lines using the following
statements:
\begin{itemize}
\item The product of expressions like (\ref{eq_ampl_z}) that depend
on non-intersecting subsets of $\{z_1,\ldots,z_n\}$ can be
represented in the form (\ref{eq_ampl_z}) too.
\item Operators $A$, $U$, and $L$ give polynomials of external
momenta.
\item If $\Phi$ is an expression of the form (\ref{eq_ampl_z}), then
$A\Phi$, $U\Phi$, $L\Phi$ can be represented in the form
(\ref{eq_ampl_z}) too. For example,
\begin{equation}\label{eq_a_noexp}
 A\left[\Pi_1 R_1 +
\ldots + \Pi_N R_N\right]=R_1 A\Pi_1 + \ldots + R_N A\Pi_N.
\end{equation}
This follows from the fact that $A\Gamma_{\mu}(p,q)$ can be
expressed through the values of $\Gamma_{\mu}(p,q)$ on the surface
$p^2=m^2$, $q=0$ and its first derivatives at these points along
directions $(p',q')$ such that $pp'=pq'=0$ (see the explicit formula
in ~\cite{kinoshita_lepton}). The first derivatives of $R_j$ at
these points along these directions is equal to 0 because of zero
first derivatives of scalar products of the external momenta.
\end{itemize}
\item By definition, put
\begin{equation}\label{eq_sw_to_feyn}
I(z_1,\ldots,z_n)=\lim_{\varepsilon\rightarrow +0} \int_0^{+\infty}
\lambda^{n-1} \check{f}_G(z_1\lambda,\ldots,z_n\lambda,\varepsilon)
d\lambda.
\end{equation}
The problem is reduced to the calculation of the integral
\begin{equation}\label{eq_feyn_param_integral}
\int_{z_1,\ldots,z_n>0}I(z_1,\ldots,z_n)\delta(z_1+\ldots+z_n-1)
dz_1\ldots dz_n.
\end{equation}
The integral (\ref{eq_sw_to_feyn}) is obtained analytically by using
the formula
$$
\int_0^{+\infty} \lambda^{D-1}e^{\lambda(ik-\varepsilon)}d\lambda =
\frac{(D-1)!}{(\varepsilon-ik)^D}.
$$
Note that we will always have $D>0$. This follows from the fact that
the terms with $D=0$ are nulled by the operator $A$ that is applied
to some subgraph. This is because the term in Feynman amplitude
corresponding to the minimal $D$ is proportional to $\gamma_{\mu}$
(see the explicit recipe for constructing $F$, $G$, $H$ in
~\cite{kinoshita_rules, zavialov, smirnov}).
\item We compute the integral (\ref{eq_feyn_param_integral}) numerically.
\end{enumerate}

\section{Justification of the method}\label{sec_correctness}

Justification of the correctness of the described subtraction procedure
consists of two parts:
\begin{enumerate}
\item Proof of the equality $a_{e,1}^{\new}=a_{e,1}$, where $a_{e,1}=\sum_{n\geq
1}\left(\frac{\alpha}{\pi}\right)^n A_1^{(2n)}$.
\item Demonstration that the subtraction procedure removes all
divergences in each Feynman graph.
\end{enumerate}

Let us consider the first part in the 2-loop case. 2-loop Feynman graphs for
electron's AMM are showed on Figure \ref{fig_2loop}. We must prove
that the application of this subtraction procedure is equivalent to the
on-shell renormalization. The on-shell renormalization can be
represented in the form that is similar to the one that was used for
description of the subtraction procedure in Section
\ref{subsec_feyn_ampl}, see Table \ref{table_2loop_onshell}. Here,
$B$ is the operator that is applied to Feynman amplitudes of
electron self-energy subgraphs for the on-shell renormalization. This
operator is defined by the following relation:
\begin{equation}\label{eq_b_def}
B\Sigma(p)=\Sigma(m)+(\hat{p}-m)(b(m^2)+2a'(m^2)+2mb'(m^2))
\end{equation}
if (\ref{eq_sigma_general}) is satisfied\footnote{By
definition, $\Sigma(m)=a(m^2)+mb(m^2)$.}.

\ctable[label=table_2loop_onshell,caption={Operator expressions for
contributions of graphs from Figure \ref{fig_2loop} to electrons's
AMM that are obtained directly by on-shell renormalization, the
differences between these expressions and expressions from Table
\ref{table_2loop}}.,pos=H]{|l|l|l|}{}{
\hline \# & operator expression & difference \\
\hline 1 & $A_G-A_GL_{abc}$ &
$(L_G-U_G)A_{abc}-A_G(L_{abc}-U_{abc})$
\\
\hline 2 & $A_G$ & $0$
\\
\hline 3 & $A_G-A_GL_{bcd}$ & $A_G(U_{abc}-L_{abc})$
\\
\hline 4 & $A_G-A_GL_{bcd}$ & $A_G(U_{abc}-L_{abc})$
\\
\hline 5 & $A_G-A_GB_{bc}$ & $A_G(U_{bc}-B_{bc})$
\\
\hline 6 & $A_G-A_GB_{bc}$ & $A_G(U_{bc}-B_{bc})$
\\
\hline 7 & $A_G-A_GU_{de}$ & $0$
\\
\hline }

As shown in this table, the contribution of the graph 1 to
$a_{e,1}-a_{e,1}^{\new}$ is equal to zero. Let us consider the
contribution of graphs 3--6 to this difference. Note that the
following statement is valid. Suppose the functions $\Sigma(p)$,
$\Gamma_{\mu}(p,q)$ and the complex number $C$ satisfy the following
conditions:
\begin{itemize}
\item (\ref{eq_sigma_general}),
(\ref{eq_gamma_general_q0});
\item the Ward identity:
$$
\Gamma_{\mu}(p,0)=-\frac{\partial \Sigma(p)}{\partial p^{\mu}};
$$
\item $U\Gamma_{\mu}=C\gamma_{\mu}$;
\end{itemize}
then $U\Sigma(p)=\Sigma(m)-C(\hat{p}-m)$. Also, if additionally
$(U-L)\Gamma_{\mu}=C_1\gamma_{\mu}$, then
$(U-B)\Sigma(p)=-C_1(\hat{p}-m)$. From this it follows that the
contribution of graphs 3--6 is equal to zero. The complete proof of
the relation $a_{e,1}^{\new}=a_{e,1}$ for any order of perturbation
is given in the full version of this paper~\cite{full_version}.

Let us consider the elimination of divergences in the 2-loop case.
Note that overall UV-divergences are removed by the operator $A$,
see ~\cite{kinoshita_infrared}. Thus, graph 2 doesn't have
divergences, all UV-divergences in graph 7 are obviously removed.
Also, some subgraphs can generate IR-divergences, see
~\cite{kinoshita_infrared}. The vertex that is incident to the
external photon line is a such subgraph in graphs 1--6. However,
these IR-divergences (''overall'') are removed by operator $A$, see
~\cite{kinoshita_infrared}. Each of graphs 3--6 has a unique
UV-divergent subgraph that doesn't coincide with the whole graph.
The UV-divergence corresponding to this subgraph is subtracted by
the counterterm with operator $U$. However, operator $U$ doesn't
generate additional IR-divergences (in contrast to operators $L$ and
$B$) because all IR-divergences in Feynman amplitudes like
(\ref{eq_gamma_general_q0}) are proportional to $p_{\mu}$ or
$\hat{p}p_{\mu}$, all IR-divergences in (\ref{eq_b_def}) are
contained in terms with $a'(m^2)$ or $b'(m^2)$. In the graph 1 the
subgraph $abc$ generates UV and IR divergences simultaneously. The
UV-divergence that corresponds to $abc$ is subtracted by the
counterterm $A_GU_{abc}$, this counterterm doesn't generate
additional IR-divergences. The IR-divergence that corresponds to the
subgraph $abc$ is subtracted by the counterterm
$(L_G-U_G)A_{abc}f_G$. This counterterm doesn't generate additional
UV-divergences. In this cases all divergences are eliminated
point-by-point, before integration\footnote{As was noted above, the
application of the given subtraction procedure for graph~1 is
equivalent to the on-shell renormalization. However, this
equivalence is not point-by-point in the Feynman-parametric
representation. In particular, the on-shell renormalization doesn't
lead to a convergent integral for graph~1.}. The point-by-point
elimination of divergences is described in detail in the full
version of this paper~\cite{full_version}.

\section{Application of the method to computation of $A_1^{(4)}$ and
$A_1^{(6)}$}\label{sec_realization}

The described above method of divergence elimination was applied to the
computation of 2-loop and 3-loop corrections to the electron AMM.
The purpose of this calculation is to check the
subtraction procedure. The D programming language
~\cite{dlang} was used. The code for the integrands was generated
automatically in the C programming language. The Feynman
gauge (\ref{eq_feynman_gauge}) and the following values of the constants:
$C_A=0$ from (\ref{eq_a_def}) and $C_U=0$ from (\ref{eq_u_vertex}) were used. The
numerical integration was performed by an adaptive Monte Carlo
method. Each Feynman graph was computed separately. Feynman graphs
that are obtained from each other by changing the direction of
electron lines were computed separately. The integration domain was
split into 620 and 5100 subdomains for computation of $A^{(4)}_1$
and $A^{(6)}_1$ respectively. The following probability
density function for Monte Carlo integration was used:
$$
C_{j(\underline{z})}\frac{\min(z_1,\ldots,z_n)^{s}}{z_1\ldots z_n},
$$
where $s\approx 0.74,$ $j(z)$ is the number of the subdomain containing
the tuple $\underline{z}=(z_1,\ldots,z_n)$, coefficients $C_j$ were
adjusted dynamically. The splitting of the integration domain into
subdomains is performed by the following rules:
\begin{itemize}
\item For each tuple $\underline{z}$ we determine a partition
of the set of indexes of $\underline{z}$ into two non-empty subsets
$A$ and $B$ such that $(\min_B z_l)/(\max_A z_l)$ is maximal. So, the
integration domain is split into $2^n-2$ pieces.
\item Each piece is split into $10$ parts. The number of a part for
the tuple $\underline{z}$ is the number of interval from the list
$$
[0;1],\ (1;2],\ (2;3],\ (3,5],\ (5,7],$$ $$(7,10],\ (10,14],\
(14,18],\ (18,22],\ (22,+\infty),
$$
that contains the value $\ln((\min_B z_l)/(\max_A z_l))$.
\end{itemize}

The integrand appears as a difference of functions such that the
corresponding integrals can diverge. Moreover, these divergences can
have a linear character or even more. Thus, round-off errors can
introduce a significant contribution to the result. In the cases
when the 64-bit precision was not enough, we used the 320-bit precision
(with the help of the SCSLib library ~\cite{scslib}). These situations appear
with the probability of about $1/2000$ during the Monte Carlo integration.
The situations when the 320-bit precision is not enough appear with
the probability less than $10^{-9}$ and don't introduce any noticeable
contribution (these points are discarded).

3 days of computation on a personal computer give the following
result:
$$
A^{(4)}_1=-0.328513(87), $$
$$
A_1^{(6)}=1.1802(85)
$$
(the uncertainties hereinafter represent the $90\%$ confidential
limits). These results are in good agreement with
(\ref{eq_analyt2}), (\ref{eq_analyt3}). The uncertainties are due to
the statistical error of the Monte Carlo integration. These
uncertainties can be made arbitrarily small by increasing the time
of computation.

\begin{figure}[H]
\includegraphics[scale=1]{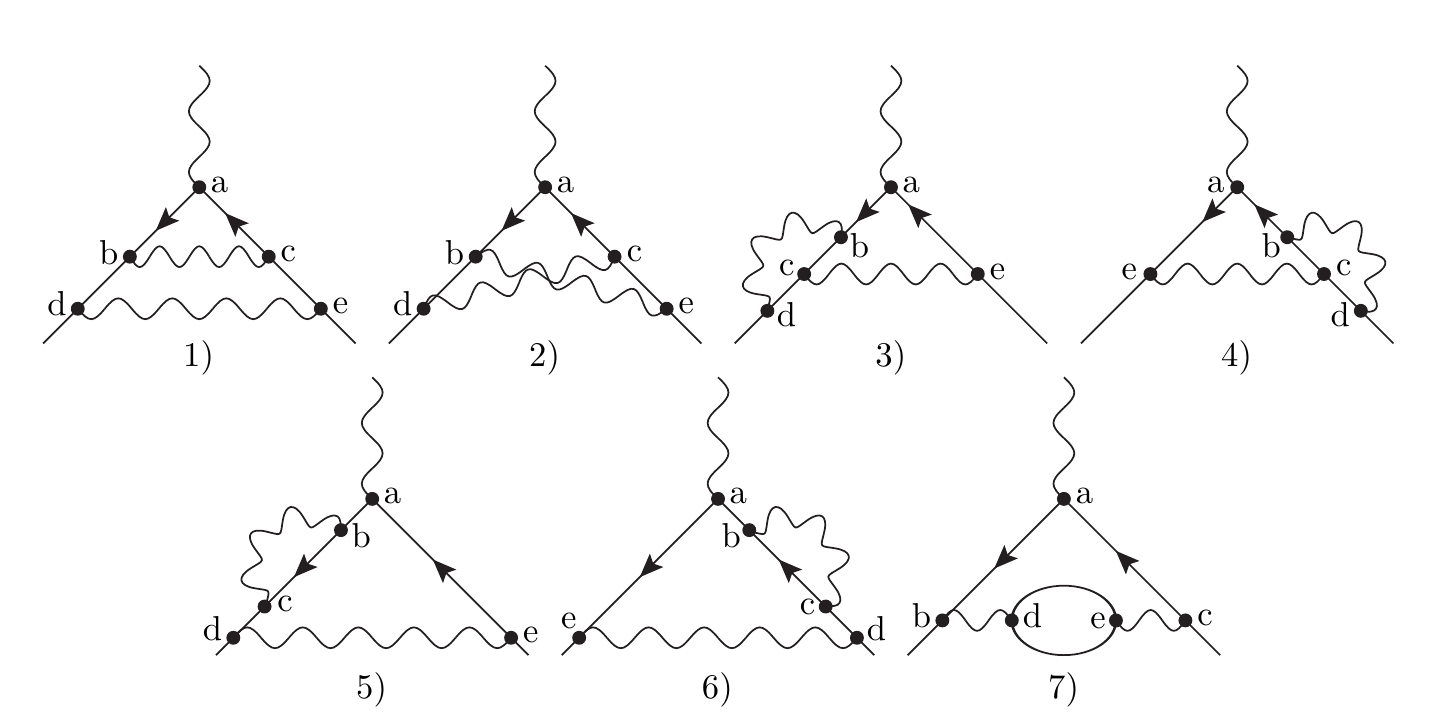}
\caption{2-loop Feynman graphs for electron's AMM.} \label{fig_2loop}
\end{figure}

7 Feynman graphs contributing to $A^{(4)}_1$ are showed on Figure
\ref{fig_2loop}, contributions of each graph to $A^{(4)}_1$ are
presented in Table \ref{table_2loop}. In this table and in the
following tables the number $N_{\text{call}}$ denotes the number of
function calls during the numerical integration. The calculated
contribution of graphs 1, 2, and 7 are in good agreement with
well-known values that were obtained from analytical expressions
($0.77747802$, $-0.46764545$, and  $0.01568742$, respectively), see
~\cite{analyt2_kk,analyt2_p,terentiev}.

\ctable[label=table_2loop,caption={Contributions of graphs from
Figure \ref{fig_2loop} to $A_1^{(4)}$ with operator expressions from
which these contributions are obtained.},pos=H]{|l|l|l|l|}{}{
\hline \# & value & $N_{\text{call}}$ & operator expression \\
\hline 1 & $0.777455(52)$ & $5\cdot 10^{9}$ &
$A_G-A_GU_{abc}-(L_G-U_G)A_{abc}$
\\
\hline 2 & $-0.467626(44)$ & $4\cdot 10^{9}$ & $A_G$
\\
\hline 3 & $-0.032023(29)$ & $2\cdot 10^{9}$ & $A_G-A_GU_{bcd}$
\\
\hline 4 & $-0.032033(29)$ & $2\cdot 10^{9}$ & $A_G-A_GU_{bcd}$
\\
\hline 5 & $-0.294978(25)$ & $2\cdot 10^{9}$ & $A_G-A_GU_{bc}$
\\
\hline 6 & $-0.294998(24)$ & $2\cdot 10^{9}$ & $A_G-A_GU_{bc}$
\\
\hline 7 & $0.0156895(25)$ & $2\cdot 10^{9}$ & $A_G-A_GU_{de}$
\\
\hline }

\begin{figure}[H]
\begin{flushleft}
\includegraphics[scale=0.65]{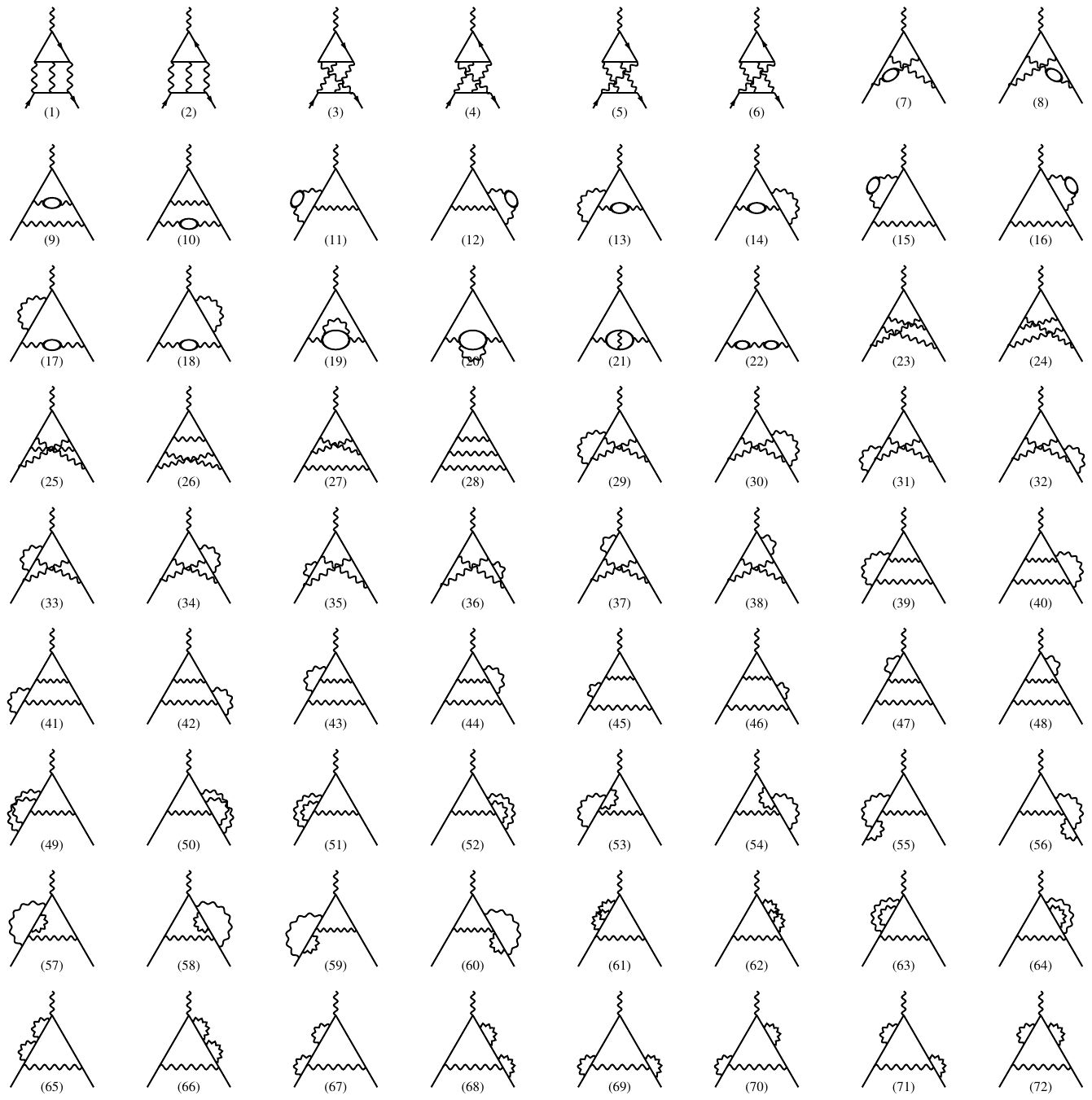}
\caption{3-loop Feynman graphs for electron's AMM. Plot courtesy of
F.~Jegerlehner }\label{fig_3loop}
\end{flushleft}
\end{figure}

72 Feynman graphs that contribute to $A^{(6)}_1$ are shown on Figure
\ref{fig_3loop}. The computed contributions of each graph to
$A^{(6)}_1$ are presented in Table \ref{table_3loop}. Table
\ref{table_3loop_cmp} contains the comparison of the computed
contributions of some sets of graphs with known values for
these contributions. We selected such sets of graphs that their
contributions calculated by the given subtraction procedure is equal (should be equal)
to the contributions computed directly in the Feynman gauge. If all
computations would be performed with 64-bit precision, the points
for which this precision is not enough would be discarded, then it
appears an additional error being more than $0.1\%$ in graphs
28, 35--36, 45--46, 70--71.

\ctable[label=table_3loop,caption={Contributions of graphs from
Figure \ref{fig_3loop} to $A_1^{(6)}$.}, pos=H]{|l|l|l|l|l|l|}{}{
\hline \# & value & $N_{\text{call}}$ & \# & value & $N_{\text{call}}$ \\
\hline 1 & $-0.17836(43)$ & $9\cdot 10^{7}$
 & 37 & $0.15782(86)$ & $10^{8}$
\\
\hline 2 & $-0.17862(43)$ & $9\cdot 10^{7}$
 & 38 & $0.15674(86)$ & $2\cdot 10^{8}$
\\
\hline 3 & $0.18220(61)$ & $10^{8}$
 & 39 & $-0.16718(67)$ & $10^{8}$
\\
\hline 4 & $0.18217(61)$ & $10^{8}$
 & 40 & $-0.16750(67)$ & $10^{8}$
\\
\hline 5 & $0.18155(61)$ & $10^{8}$
 & 41 & $-1.6283(17)$ & $3\cdot 10^{8}$
\\
\hline 6 & $0.18188(62)$ & $10^{8}$
 & 42 & $-1.6283(17)$ & $3\cdot 10^{8}$
\\
\hline 7 & $-0.042127(73)$ & $7\cdot 10^{7}$
 & 43 & $0.3054(10)$ & $2\cdot 10^{8}$
\\
\hline 8 & $-0.042245(74)$ & $7\cdot 10^{7}$
 & 44 & $0.3058(10)$ & $2\cdot 10^{8}$
\\
\hline 9 & $0.11513(16)$ & $8\cdot 10^{7}$
 & 45 & $2.2553(19)$ & $4\cdot 10^{8}$
\\
\hline 10 & $0.019128(59)$ & $7\cdot 10^{7}$
 & 46 & $2.2569(18)$ & $4\cdot 10^{8}$
\\
\hline 11 & $0.028301(50)$ & $7\cdot 10^{7}$
 & 47 & $-1.1357(11)$ & $2\cdot 10^{8}$
\\
\hline 12 & $0.028278(51)$ & $7\cdot 10^{7}$
 & 48 & $-1.1341(11)$ & $2\cdot 10^{8}$
\\
\hline 13 & $-0.015016(72)$ & $7\cdot 10^{7}$
 & 49 & $0.10755(42)$ & $10^{8}$
\\
\hline 14 & $-0.015153(72)$ & $7\cdot 10^{7}$
 & 50 & $0.10718(41)$ & $10^{8}$
\\
\hline 15 & $-0.072255(92)$ & $7\cdot 10^{7}$
 & 51 & $-0.01470(38)$ & $9\cdot 10^{7}$
\\
\hline 16 & $-0.072144(93)$ & $7\cdot 10^{7}$
 & 52 & $-0.01421(38)$ & $9\cdot 10^{7}$
\\
\hline 17 & $-0.041038(94)$ & $7\cdot 10^{7}$
 & 53 & $0.07232(57)$ & $10^{8}$
\\
\hline 18 & $-0.041091(93)$ & $7\cdot 10^{7}$
 & 54 & $0.07212(56)$ & $10^{8}$
\\
\hline 19 & $0.019872(72)$ & $7\cdot 10^{7}$
 & 55 & $-0.8390(10)$ & $2\cdot 10^{8}$
\\
\hline 20 & $0.019857(71)$ & $7\cdot 10^{7}$
 & 56 & $-0.8394(10)$ & $2\cdot 10^{8}$
\\
\hline 21 & $0.013153(88)$ & $7\cdot 10^{7}$
 & 57 & $0.40154(69)$ & $10^{8}$
\\
\hline 22 & $0.002548(20)$ & $6\cdot 10^{7}$
 & 58 & $0.40295(69)$ & $10^{8}$
\\
\hline 23 & $0.9311(10)$ & $2\cdot 10^{8}$
 & 59 & $0.41612(82)$ & $10^{8}$
\\
\hline 24 & $0.9318(10)$ & $2\cdot 10^{8}$
 & 60 & $0.41581(81)$ & $10^{8}$
\\
\hline 25 & $-0.02688(47)$ & $10^{8}$
 & 61 & $1.2625(11)$ & $2\cdot 10^{8}$
\\
\hline 26 & $-0.9458(11)$ & $2\cdot 10^{8}$
 & 62 & $1.2620(11)$ & $2\cdot 10^{8}$
\\
\hline 27 & $-2.2306(19)$ & $4\cdot 10^{8}$
 & 63 & $-0.02913(63)$ & $10^{8}$
\\
\hline 28 & $1.7888(19)$ & $4\cdot 10^{8}$
 & 64 & $-0.02911(62)$ & $10^{8}$
\\
\hline 29 & $-0.87900(74)$ & $10^{8}$
 & 65 & $-1.0614(10)$ & $2\cdot 10^{8}$
\\
\hline 30 & $-0.87894(74)$ & $10^{8}$
 & 66 & $-1.0620(10)$ & $2\cdot 10^{8}$
\\
\hline 31 & $2.5206(17)$ & $3\cdot 10^{8}$
 & 67 & $-0.04893(73)$ & $10^{8}$
\\
\hline 32 & $2.5207(17)$ & $3\cdot 10^{8}$
 & 68 & $-0.04831(72)$ & $10^{8}$
\\
\hline 33 & $0.07018(50)$ & $10^{8}$
 & 69 & $-2.9084(20)$ & $4\cdot 10^{8}$
\\
\hline 34 & $0.07017(50)$ & $10^{8}$
 & 70 & $3.2668(21)$ & $5\cdot 10^{8}$
\\
\hline 35 & $-1.7479(16)$ & $3\cdot 10^{8}$
 & 71 & $3.2652(21)$ & $4\cdot 10^{8}$
\\
\hline 36 & $-1.7498(16)$ & $3\cdot 10^{8}$
 & 72 & $-3.2047(20)$ & $4\cdot 10^{8}$
\\
\hline }

\ctable[caption={Comparison of contributions of some sets of Feynman
graphs from Figure \ref{fig_3loop} to $A_1^{(6)}$ with known
analytical values.}, label=table_3loop_cmp,
pos=H]{|l|l|l|l|}{\tnote[a] {There is a disagreement of
contributions of sets 7--8 and 9--10 separately with values that are
given in that papers. We failed to find the reason, but the
contribution of 7--10 is in good agreement,  the error is less
than $0.6\%$.}}{
\hline \# & value & analyt. value & Ref. \\
\hline 1--6 & $0.3708(14)$ & $0.3710052921$ & ~\cite{analyt_ll}
\\
\hline 7--10 & $0.04989(20)$ & $0.05015$ &
~\cite{analyt_b1,analyt_b4}\tmark[a]
\\
\hline 11--12,15--16 & $-0.08782(15)$ & $-0.0879847$ &
~\cite{analyt_b2,analyt_b1}
\\
\hline 13--14,17--18 & $-0.11230(17)$ & $-0.112336$ &
~\cite{analyt_b3,analyt_b1}
\\
\hline 19--21 & $0.05288(13)$ & $0.05287$ & ~\cite{analyt_mi}
\\
\hline 22 & $0.002548(20)$ & $0.0025585$ & ~\cite{analyt_mi}
\\
\hline 23--24 & $1.8629(14)$ & $1.861907872591$ & ~\cite{analyt_f}
\\
\hline 25 & $-0.02688(47)$ & $-0.026799490$ & ~\cite{analyt3}
\\
\hline 26--27 & $-3.1764(22)$ & $-3.17668477$ & ~\cite{analyt_d}
\\
\hline 28 & $1.7888(19)$ & $1.79027778$ & ~\cite{analyt_d}
\\
\hline 29--30 & $-1.7579(10)$ & $-1.757936342$ & ~\cite{analyt3}
\\
\hline 31--32,37--38 & $5.3559(27)$ & $5.35763265$ &
~\cite{analyt_d,analyt_c}
\\
\hline 33--34,37--38 & $0.4549(14)$ & $0.45545185$ &
~\cite{analyt_d,analyt_f}
\\
\hline 31--32,35--36 & $1.5436(34)$ & $1.541649$ &
~\cite{analyt_e,analyt_c}
\\
\hline 33--36 & $-3.3573(24)$ & $-3.360532$ &
~\cite{analyt_e,analyt_f}
\\
\hline 39--40 & $-0.33468(95)$ & $-0.334695103723$ &
~\cite{analyt_f}
\\
\hline 41--48 & $-0.4030(41)$ & $-0.4029$ &
~\cite{analyt_b,analyt_e}
\\
\hline 49--72 & $0.9529(53)$ & $0.9541$ &
~\cite{analyt_b,analyt_e,analyt_d,analyt_c,analyt_f,analyt3}
\\
\hline }

\section{Acknowledgments}

I am grateful to A.L.~Kataev for many useful discussions and
recommendations, help in organizational issues, to O.V.~Teryaev for
helpful discussions and help in organizational issues, to
L.V.~Kalinovskaya for help in organizational issues, to A.B.~Arbuzov
for help in preparing this version of the text and help in
organizational issues.

This research was partially supported by RFBR Grant N 14-01-00647.


\begin{thebibliography}{99}
\bibitem{bogolubovparasuk} N.N. Bogoliubov, O.S. Parasiuk // Acta Math. 97, 227 (1957).
\bibitem{hepp} K. Hepp Proof of the Bogoliubov-Parasiuk Theorem on
Renormalization // Commun. math. Phys. --- 1966. --- V.2. ---
301--326.
\bibitem{scherbina} V.A. Scherbina // Catalogue of Deposited Papers, VINITI, Moscow,
 38, 1964 (in Russian).
\bibitem{zavialovstepanov} O.I. Zavialov, B.M. Stepanov // Yadernaja Fysika
(Nuclear Physics) 1, 922, 1965 (in Russian).
\bibitem{zimmerman} W. Zimmermann, Convergence of Bogoliubov's Method
of Renormalization in Momentum Space // Commun. math. Phys. ---
1969. --- V. 15. --- 208--234.
\bibitem{ll4} V.B. Berestetskii, E.M. Lifshitz, L.P. Pitaevskii,
Quantum Electrodynamics, Butterworth-Heinemann, 1982.
\bibitem{bogoliubov_shirkov} N.N. Bogoliubov, D.V. Shirkov, Introduction to
the Theory of Quantized Fields, John Wiley \& Sons Inc, 1980.
\bibitem{experiment} D. Hanneke, S. Fogwell Hoogerheide,
G. Gabrielse, Cavity control of a single-electron quantum cyclotron:
Measuring the electron magnetic moment // Physical Review A. ---
2011. --- V. 83, 052122.
\bibitem{kinoshita_10} T. Aoyama, M. Hayakawa, T. Kinoshita, M. Nio,
Tenth-Order QED Contribution to the Electron g - 2 and an Improved
Value of the Fine Structure Constant // Physical Review Letters
--- 2012. --- V. 109, 111807.
\bibitem{kinoshita_10_new} T. Aoyama, M. Hayakawa, T. Kinoshita, M.
Nio, Tenth-Order Electron Anomalous Magnetic Moment -- Contribution
of Diagrams without Closed Lepton Loops // Physical Review D. ---
2015.
--- V. 91, 033006.
\bibitem{schwinger1} J. Schwinger, On Quantum Electrodynamics and the
magnetic moment of the electron // Physical Review. --- 1948. --- V.
73. --- 416.
\bibitem{schwinger2} J. Schwinger, Quantum Electrodynamics, III: the
electromagnetic properties of the electron --- radiative corrections
to scattering // Physical Review. --- 1949. --- V. 76. --- 790.
\bibitem{analyt2_kk} R. Karplus, N. Kroll, Fourth-order corrections
in Quantum Electrodynamics and the magnetic moment of the electron
// Physical Review. --- 1950. --- V. 77, N. 4. --- 536--549.
\bibitem{analyt2_p} A. Petermann, Fourth order magnetic moment of the
electron // Helvetica Physica Acta. --- 1957. --- V. 30.
--- 407--408.
\bibitem{analyt2_z} C. Sommerfield, Magnetic dipole moment of the electron //
Physical Review. --- 1957. -- N. 107. --- 328--329.
\bibitem{terentiev} M.V. Terentiev // Soviet Physics JETP, 16, 444
(1963).
\bibitem{levinewright} M. Levine, J. Wright, Anomalous magnetic
moment of the electron // Physical Review D. --- 1973. --- V. 8, N.
9. --- 3171--3180.
\bibitem{carrollyao} R. Carroll, Y. Yao, $\alpha^3$ contributions to
the anomalous magnetic moment of an electron in the mass-operator
formalism // Physics Letters. --- 1974. --- V. 48B, N. 2. ---
125--127.
\bibitem{carroll} R. Carroll, Mass-operator calculation of the
electron $g$ factor // Physical Review D. --- 1975. --- V. 12, N. 8.
--- 2344--2355.
\bibitem{kinoshita_6} P. Cvitanovi\'{c}, T. Kinoshita,
Sixth-order magnetic moment of the electron // Physical Review D.
--- 1974. --- V. 10, N. 12.
--- pp. 4007--4031.
\bibitem{kinoshita_6_prec} T. Kinoshita, New Value of the $\alpha^3$
electron anomalous magnetic moment // Physical Review Letters. ---
1995. --- V. 75, N. 26. --- 4728--4731.
\bibitem{analyt_mi} J. Mignaco, E. Remiddi, Fourth-order vacuum
polarization contribution to the sixth-order electron magnetic
moment // IL Nuovo Cimento. --- 1969. --- V. LX A, N. 4. ---
519--529.
\bibitem{analyt_b2} R. Barbieri, M. Caffo, E. Remiddi, A contribution
to sixth-order electron and muon anomalies. -- II // Lettere al
Nuovo Cimento. --- 1972. -- V. 5, N. 11. --- 769--773.
\bibitem{analyt_b3} D. Billi, M. Caffo, E. Remiddi, A Contribution to the sixth-Order
electron and muon Anomalies // Lettere al Nuovo Cimento. --- 1972.
--- V. 4, N. 14. --- 657--660.
\bibitem{analyt_b1} R. Barbieri, E. Remiddi, Sixth order electron and
muon $(g-2)/2$ from second order vacuum polarization insertion //
Physics Letters. --- 1974. --- V. 49B, N. 5. --- 468--470.
\bibitem{analyt_b4} R. Barbieri, M. Caffo, E. Remiddi, A contribution
to sixth-order electron and muon anomalies -- III //
Ref.TH.1802-CERN. --- 1974.
\bibitem{analyt_b} M. Levine, R. Roskies, Hyperspherical approach to quantum electrodynamics: sixth-order magnetic
moment // Physical Review D. --- 1974. --- V. 9, N. 2. --- 421--429.
\bibitem{analyt_e} M. Levine, R. Perisho, R. Roskies, Analytic
contributions to the $g$ factor of the electron // Physical Review
D. --- 1976. --- V. 13, N. 4. --- 997--1002.
\bibitem{analyt_d} R. Barbieri, M. Caffo, E. Remiddi, S. Turrini,
D. Oury, The anomalous magnetic moment of the electron in QED: some
more sixth order contributions in the dispersive approach // Nuclear
Physics B. --- 1978. --- N. 144. --- 329--348.
\bibitem{analyt_c} M. Levine, E. Remiddi, R. Roskies, Analytic
contributions to the $g$ factor of the electron in sixth order //
Physical Review D. --- 1979. --- V. 20, N. 8. --- 2068--2077.
\bibitem{analyt_ll} S. Laporta, E. Remiddi, The analytic value of
the light-light vertex graph contributions to the electron $g-2$ in
QED // Physics Letters B. --- 1991. --- N. 265. --- 182--184.
\bibitem{analyt_f} S. Laporta, The analytical value of the
corner-ladder graphs contribution to the electron $(g-2)$ in QED //
Physics Letters B. --- 1995. --- N. 343. --- 421--426.
\bibitem{analyt3} S. Laporta, E. Remiddi, The Analytical value of the
electron (g-2) at order $\alpha^3$ in QED // Physical Letters B. ---
1996. --- V. 379. --- 283--291.
\bibitem{kinoshita_8start} T. Kinoshita, W. Lindquist, Eighth-order
anomalous magnetic moment of the electron // Physical Review
Letters. --- 1981. --- V. 47, N. 22. --- 1573--1576.
\bibitem{kinoshita_watanabe} T. Aoyama, M. Hayakawa, T. Kinoshita,
M. Nio, N. Watanabe, Eighth-order vacuum-polarization function
formed by two light-by-light-scattering diagrams and its
contribution to the tenth-order electron $g-2$ // Physical Review D.
--- 2008. --- N. 78., 053005.
\bibitem{rubidium} R. Bouchendira, P. Clad\'{e}, S. Guellati-Kh\'{e}lifa,
F. Nez, F. Biraben, New Determination of the Fine Structure Constant
and Test of the Quantum Electrodynamics // Physical Review Letters.
--- 2011. --- V. 106, 080801.
\bibitem{codata} P. Mohr, B. Taylor, D. Newell, CODATA recommended values of the fundamental physical constants:
2010* // Reviews of Modern Physics. --- 2012. --- V.84, 1527.
\bibitem{analytheavy4_kataev} A.L. Kataev, Analytical eighth-order
light-by-light QED contributions from leptons with heavier masses to
the anomalous magnetic moment of the electron // Physical Review D.
--- 2012. --- N. 86, 013010.
\bibitem{analytheavy4} A. Kurz, T. Liu, P. Marquard, M. Steinhauser,
Anomalous magnetic moment with heavy virtual leptons // Nuclear
Physics B. --- 2014. --- V. 879. --- 1--18.
\bibitem{full_version} S. Volkov, Subtractive procedure for calculating the anomalous electron
magnetic moment in QED and its application for numerical calculation
at the three-loop level, J. Exp. Theor. Phys. (2016), V. 122, N. 6,
pp. 1008--1031.
\bibitem{zavialov} O.I. Zavialov, Renormalized Quantum Field Theory,
Springer Science \& Business Media, 2012.
\bibitem{smirnov} V.A. Smirnov, Renormalization and Asymptotic Expansions,
PPH'14 (Progress in Mathematical Physics), Birkh\"{a}user, 2000.
\bibitem{kinoshita_rules} P. Cvitanovi\'{c}, T. Kinoshita,
Feynman-Dyson rules in parametric space // Physical Review D. ---
1974. --- V. 10, N. 12. --- pp. 3978--3991.
\bibitem{kinoshita_infrared} P. Cvitanovi\'{c}, T. Kinoshita,
New approach to the separation of ultraviolet and infrared
divergences of Feynman-parametric integrals // Physical Review D.
--- 1974. --- V. 10, N. 12.
--- pp. 3991--4006.
\bibitem{kinoshita_lepton} T. Aoyama, M. Hayakawa, T. Kinoshita, M.
Nio, Automated calculation scheme for $\alpha^n$ contributions of
QED to lepton g-2: Generating renormalized amplitudes for diagrams
without lepton loops // Nuclear Physics B. --- 2006. --- V. 740. ---
pp. 138--180.
\bibitem{dlang} A. Alexandrescu, The D Programming Language,
Addison-Wesley Professional, 2010.
\bibitem{scslib} D. Defour, F. Dinechin, Software Carry-Save for fast
multiple-precision algorithms // Proceedings of the First
International Congress of Mathematical Software, Beijing, China,
2002.

\end{thebibliography}
\end{document}